\documentclass[pdflatex,sn-vancouver-ay]{sn-jnl}
\usepackage{graphicx}
\usepackage{multirow}
\usepackage{xcolor}
\usepackage{textcomp}
\usepackage{amsmath}
\usepackage{longtable}
\usepackage{geometry}
\usepackage{array}

\title{\textbf{The promise and perils of AI in medicine}}

\author[1]{Robert Sparrow}

\author[2]{Joshua Hatherley}

\affil[1]{School of Philosophical, Historical, and International Studies, Monash University, Australia}

\affil[2]{Center for the Philosophy of AI, University of Copenhagen, Denmark}

\abstract{What does Artificial Intelligence (AI) have to contribute to health care? And what should we be looking out for if we are worried about its risks? In this paper we offer a survey, and initial evaluation, of hopes and fears about the applications of artificial intelligence in medicine. AI clearly has enormous potential as a research tool, in genomics and public health especially, as well as a diagnostic aid. It’s also highly likely to impact on the organisational and business practices of healthcare systems in ways that are perhaps under-appreciated. Enthusiasts for AI have held out the prospect that it will free physicians up to spend more time attending to what really matters to them and their patients. We will argue that this claim depends upon implausible assumptions about the institutional and economic imperatives operating in contemporary healthcare settings. We will also highlight important concerns about privacy, surveillance, and bias in big data, as well as the risks of over trust in machines, the challenges of transparency, the deskilling of healthcare practitioners, the way AI reframes healthcare, and the implications of AI for the distribution of power in healthcare institutions. We will suggest that two questions, in particular, are deserving of further attention from philosophers and bioethicists. What does care look like when one is dealing with data as much as people? And, what weight should we give to the advice of machines in our own deliberations about medical decisions?

\bigskip

This is a pre-print of: Sparrow, Robert, and Joshua Hatherley. 2019. The promise and perils of AI in medicine. \textit{International Journal of Chinese and Comparative Philosophy of Medicine}, 17(2): 79-109. \href{http://dx.doi.org/10.24112/ijccpm.171678}{10.24112/ijccpm.171678}}

\begin{document}

\maketitle

\section{Introduction}

AI encompasses a vast range of diverse technologies, and AI research cuts across a variety of disciplines including computer science, philosophy of mind, logic, neuroscience, and theoretical biology. As such, a precise definition is difficult. For our purposes, though, it will suit to suggest that AI deals with the creation of machines capable of acting rationally or intelligently \citep{russell2016artificial}. 

Famously, the initiators of the AI research program in the mid-1950s thought that it would only take a few months to bear fruit. After some early successes, the 1970s and 80s saw a significant drop in funding and excitement about AI – the so-called AI winter. However, recent advances in machine learning, and especially “deep learning”, prompted by the increase in available computing power and emergence of large datasets as a result of the Internet, have led to its re-emergence into the foreground of public awareness: an ‘AI spring’ is now blooming. Machine learning involves the creation of machines able to learn (semi-)autonomously from experience. Deep learning is a type of machine learning technique that uses so-called ‘deep’ neural networks (i.e. networks consisting of multiple hidden layers) to generate impressively accurate predictions and classifications. These networks are inspired by the neural architecture of the human brain, in that they consist of a complex network of interconnected ‘neurons’. An important application of deep learning, especially relevant to the applications of AI in medicine, is natural language processing, which involves the creation of models able to identify, process, and perform actions in response to written text and/or to speech. 

Medicine is one of the areas where there is the most enthusiasm about the application of AI. There are at least four reasons for this. First, AI is heavily reliant upon the availability of large-scale, varied datasets – so-called ‘big’ data. The digitisation of modern healthcare has generated an enormous amount of data that AI can take advantage of. Data currently being stored in electronic health records, for instance, is expected to be the first port of call for large scale developments and applications of medical AI, in addition to the data that has proliferated from wearables (e.g. FitBits), online patient forums (e.g. PatientsLikeMe), even credit card transactions \citep{weber2014finding}.  Second, people care about — and are willing to spend money on — their health, meaning that there is money to be made. Technology corporations, small and large, are hoping to capitalise on this new domain of healthcare technology, and so are investing heavily. Third, by improving health, AI has significant potential to help people, so there are many good people passionately dedicating themselves to this end. Finally, governments are very concerned about the size of their healthcare budgets and hope that, by identifying which treatments work and which don’t as well as by discovering new drugs and new treatments, AI might help them reduce the cost of healthcare.\footnote{Unfortunately, the economics of healthcare suggests that technological advances in medicine actually contribute to, rather than reduce, the total cost of healthcare, mostly because the total amount governments spend on each individual only increases with life expectancy, as older people have more complex medical needs \citep{callahan2009taming}.}

\section{The promise of AI in medicine}

Our purposes here are inevitably, for the most part, critical, in the service of what we hope is the laudable goal of drawing attention to ethical and political problems that need to be addressed in order to maximise the benefits and minimise the risks from the application of AI in healthcare. It is therefore important that we clearly state at the outset our belief that AI does have tremendous promise when it comes to the goals of medicine. In particular, we anticipate that AI will ultimately produce significant benefits in the area of research, diagnosis, and medical administration. In this section, we also consider its potential to “rehumanise” medicine by facilitating more and better communication between physicians and patients.

\subsection{AI for research}

Increasingly, research in biology and medicine involve generating, manipulating, and analysing large datasets. One of AI’s primary strengths is its ability to identify patterns in data, where these might escape human beings, and for this reason AI has extraordinary potential as a tool for medical research. 

AI is already being applied in research application in genomics, drug discovery and design, and to data mine EHR systems to identify novel and clinically useful phenotypes and biomarkers of illness. Research on genomics is highly reliant on big data and sophisticated algorithms play a crucial role in genetic sequencing as well as in genome wide association studies. AI is being used to assist in the development of new drug compounds, through the prediction and identification of potentially productive and efficacious molecules. For instance, Deep Mind’s “Alphafold” AI uses machine learning to generate models of proteins based on their genetic sequences \citep{alquraishi2019alphafold}. Additionally, AI has been applied for the purpose of data mining EHR systems to discover information relevant to individual patients as well as information about populations \citep{chen2017textual}. There is also significant potential for AI to use novel sources of data, including that generated by mobile devices and individuals’ activity online, to generate new findings in public health.

The uptake of AI for medical research has been more rapid than in clinical practice, due the more demanding regulatory regime that governs clinical applications of new technologies.  The relative lack of regulatory oversight of research is one reason to be somewhat cautious about the claims made on behalf of AI.  Another is the fact that, owing to the rate of rapid progress in the area, many of these claims have been made on the basis of papers that have appeared on prepress servers rather than in the refereed literature. It is also worth noting that research using AI often raises significant ethical issues relating to consent to the use of data that have not always been handled well \citep{kahn2019alphabet}. Finally, it is important to acknowledge here that, despite its treatment in the popular press, AI isn’t magic. The findings based upon AI still rely on having good sources of data, a good understanding of data, good understanding of causal relations, and good experimental design.\footnote{This is especially important given that the reasoning of most AI systems does not factor in causation: their results are based exclusively upon correlations contained within a dataset.} Human error in the interpretation of an AI system’s result can lead even the most sophisticated AI astray. These reasons for caution when it comes to some of the less critical claims about AI are, however, no reason to deny the general claim that AI is likely to contribute greatly to medical research over the next several decades.

\subsection{AI for diagnosis}

Barely a week goes by without some new announcement about AI outperforming human physicians in some diagnostic task. In particular, ‘deep learning’ has shown significant potential for diagnosis in the context of medical imaging. For instance, there have been some promising results in the use of AI in diagnosing diabetic retinopathy \citep{gulshan2016development}, skin cancers \citep{esteva2017dermatologist}, and breast cancer \citep{golden2017deep}, among others. Consequently, medical disciplines that are heavily reliant upon the analysis of medical images are frequently considered to be most likely to be disrupted by the deployment of medical AI – dermatology, radiology, pathology, etc. Indeed, attempts are already being made to reorient these professions toward modified roles and duties that enable a productive relationship between AI systems and clinicians \citep{jha2016adapting}. 

Attempts have also been made to apply AI to diagnosis and treatment recommendation outside of medical imaging. A number of researchers are developing AI systems to work with clinical data produced by ECGs or medical monitoring devices used in ICUs in order to predict the future trajectory of patients’ conditions: in many ways such systems do not seem all that different to the kind of algorithmic medicine that has been practised in ICUs for the last several decades. However, some teams have much more ambitious agendas for AI. Notoriously, for instance, IBM’s Watson utilises both machine learning and natural language processing to trawl through the medical literature in order to better recommend treatments for patients \citep{somashekhar2017abstract}. While initial reporting about Watson tended to be wildly enthusiastic, it is fair to say that subsequent commentary has been more mixed \citep{strickland2018layoffs}. Nevertheless, Watson is undoubtedly significant as an indicator of the scope of the ambitions that AI researchers have when it comes to the role that AI might play in medicine in the future. Indeed, some pundits are now imagining that AI will finally realise the long-heralded potential of “personalised medicine” by analysing the patient’s entire genome, as well as multiple lifestyle factors, before recommending treatment for their condition \citep{topol2019deep}.

It is important to note, though, that there is more excitement about the potential of AI for diagnosis than actual clinical application. Most studies comparing the performance of clinicians and AI systems in diagnostic tasks have suffered from significant methodological limitations \citep{liu2019comparison}. As such, few of the promising results have been clinically validated. It’s one thing for an AI to replicate or exceed the performance of human beings at some classification task after access to a properly labelled dataset. It’s another for an AI to go transistor to toe with human beings in medical diagnosis as it happens in reality. 

\subsection{AI for admin}

AI also has lots of promise in an area that is less glamorous: medical administration. 

Computers and expert systems already play an important role in the complex scheduling tasks that are central to modern hospitals and healthcare systems, as well as in purchasing and billing systems, and medical data management more generally. AI will enable institutions to automate more of these business processes and to increase their efficiency. In the near future, AI systems will bill patients, roster staff, manage inventories, monitor employee performance, and schedule surgeries. As natural language interfaces improve, patients first contact with medical institutions may well be with an AI, which will make appointments for them, or direct them to the appropriate person to assist them with their enquiries.

Another, related, area where AI looks set to have a large impact is in the insurance industry. The business model of insurance industry is based around insurers being better able to identify, quantify, and manage risks than their clients. Because the large insurance agencies insure — and process claims from — millions of customers, they have correspondingly large datasets. If they can leverage AI to gain an improved understanding of risks and risk profiles, they will be able to improve their market share and/or profit margins by offering lower premiums. Of course, it is also possible that insurers might discover that it simply isn’t worth offering insurance to certain customers or classes of customers. In order to secure insurance, people with high medical needs, or poor risk profiles must be able to pool their risk with a larger group of people with lower risk profiles. However, the more accurately insurers are able to estimate risk, the more they are able to distinguish between different pools. Some individuals with complex medical conditions, or who are otherwise at high risk of requiring expensive treatment, may eventually find that the pools in which they are placed are small and include only other patients with similar risk profiles, with the result that they are unable to afford the premiums available to them \citep{price2019privacy}.

AI is also likely to play a role in Managed Care. With more data on patients and the success rates of various treatments, managed care organisations will be better placed to estimate both the likelihood and probable extent of benefits to particular patients from particular treatments and also the cost of providing them. AIs may effectively become the gatekeepers that determine who gets access to what care, when, and for how long. Indeed, there is evidence that this is already taking place \citep{lecher2018happens}.

Patients are unlikely to be terribly enthusiastic about the applications of AI in these domains. Notwithstanding the amount of effort that is going into providing machines with emotional intelligence, the experience of dealing with AI is likely to be an impersonal one, and perhaps an alienating one. Even if AI systems are more efficient than existing telephone queues and institutional bureaucracies, patients are unlikely to feel empowered by their interactions with them. As we discuss further below, in some cases, they may well be disempowered. As we also discuss further below, the use of “black box” systems to determine who gets access to healthcare also raises questions about procedural justice and respect for persons. That being said, patients will also benefit from a more efficient allocation of healthcare resources as a result of the use of AI in medical administration.

\subsection{AI for people}

It is by now widely acknowledged by clinicians that patients as individual persons have drifted to the periphery of medicine over the last half-century \citep{cassell2002doctoring, gawande2014being, topol2019deep}. Treating the disease or disorder has now become the primary focus. Many plausible explanations have been given for this phenomena: the rise of managed care \citep{mechanic1996impact}, the creation of large and impersonal medical institutions, the various conflicts of interests that have been introduced into the doctor-patient relationship \citep{rodwin1995medicine}. EHR systems have been an especially detrimental addition to the doctor-patient relationship, impeding communication \citep{toll2012cost}, increasing administrative burden \citep{hill20134000}, and contributing to physician burnout and depression \citep{friedberg2014factors}. \cite{verghese2008culture} complains that medicine is now more concerned with treating the iPatient – the digitised collection of scans, documents, and data – than the individual flesh-and-blood patient.

A number of authors have suggested that AI has significant potential to counter this trend and make the practice of medicine “more human” \citep{israni2019humanizing, mesko2018will, topol2019deep}. In particular, they suggest, automating the input and retrieval of patient data with AI might allow clinicians to return the patient to the centre of their attention.  ‘One of [AI’s] most important effects’, claims Eric Topol, ‘will come from unshackling clinicians from electronic health records’ \citep[288]{topol2019deep}. 

It is not entirely clear what this group of thinkers anticipates that machines will be doing to relieve physicians of the demands of data. Given the reliance of AI systems upon enormous datasets, one might think that advances in AI will only generate further demands on physicians when it comes to their interactions with IT systems \citep{maddox2019questions}. Indeed, the introduction of AI will itself generate an incentive to measure and collect more data, especially given that physicians will presumably need to monitor the performance of AI systems and also the outcomes for patients from the use of AI \citep{verghese2018computer}. In order to reduce the burdens described above, AI systems would have to be capable of gathering data without making further demands on human beings to respond to or manipulate it. 

It is possible that natural language processing will become advanced enough that a machine could take notes of verbal doctor-patient interactions and \textit{perhaps} even extract out those elements that are most clinically relevant. However, it’s hard to imagine that physicians would not have to at least look over these transcripts to ensure that significant information has not been missed. It’s also possible that “virtual clinical assistants” might trawl through the patient’s data, and also the relevant medical literature – as is the goal of Watson — and draw the attention of the physician to only that information that is medically relevant. This would be an ambitious application of AI and would raise many of the issues we discussed below to a large degree. Moreover, again, unless physicians were willing to concede to becoming handmaidens of diagnostic AI’s, it seems that they would need to confirm the AI’s decision through their own deliberations. It’s also likely that concerns about the legal liability of the manufacturer of the AI would lead to such AI being programmed to err on the side of inclusivity in such searches and, similarly, to physicians being required to read everything that the machine flagged as possibly relevant. Perversely, then, such systems might actually require doctors to look at more rather than less data.

Another reason to be sceptical that AI will increase the amount of time physicians have to spend time with patients derives from the economics of the provision of healthcare. Once prevention of disease and illness starts to be conceived of as part of the role of medicine, the demand for healthcare is near infinite. There is no guarantee that hospitals and other healthcare organisations will not simply take advantage of whatever efficiencies are generated by AI to move more patients through the system instead of allowing physicians to spend more time per patient. Indeed, given that patient ‘care’ is subtle and hard to measure, in contrast to the easily quantifiable amount of patients treated or procedures performed, there is every reason to think that institutions may tend to do precisely this. At this stage, then, the idea that AI will re-humanise medicine remains a commendable ambition rather than a reliable forecast. Realising this ambition will require both clever design of AI and a concerted campaign by the medical profession to resist the economic and institutional imperatives that might otherwise lead to the benefits of AI accruing primarily to institutions at the expense of the experience of doctors and patients.

\section{The perils of AI}

If AI holds out the prospect of enormous benefits, it also involves significant risks, which we survey and evaluate below. After first discussing four important issues, which have already received a significant amount of attention in the literature — privacy, surveillance, bias in big data, and “explainability” — we then move on to consider a number of issues that have received less attention to date, including:  AI’s impact on trust in medicine; its potential to deskill physicians; the danger that AI might render healthcare systems more fragile by introducing single points of failure; the likely impacts of AI on the distribution of power within healthcare institutions and systems; the vexed question of responsibility for decisions involving AI; the way in which AI may reframe healthcare; the future of care in AI-enhanced medicine; and, the enduring importance of arguments about values for the future of healthcare.

\subsection{Privacy}

Ensuring the privacy of sensitive medical information has become an increasingly challenging affair in the digital age of medicine. Digital technologies (e.g. EHR systems) have been instrumental in the effort to make patient information more easily accessible for both physicians and patients themselves. Unlike records stored on paper, electronic data is easily, and infinitely, reproducible, which makes it more accessible to non-healthcare organisations (e.g. governments, employers, and insurance agencies). Strong limitations that existed for those looking to gain access to paper records have often been weakened in the course of the adoption of digital storage of data. Additionally, electronic data is vulnerable to remote access and manipulation and thus to theft. Cyberattacks on medical organisations are becoming increasingly common due to a variety of economic incentives \citep{kruse2017cybersecurity}.

These sorts of breaches of medical confidentiality can harm patients in a number of ways. Patients may feel embarrassed, or ashamed by the idea that someone knows about their sensitive medical condition(s). They can cause patients with certain stigmatised illnesses to be alienated from their communities. They can reduce a person’s opportunities for employment. They can even lead to unwarranted increases in health insurance costs. More fundamentally, insofar as privacy is concerned with \textit{control} over information, patients are harmed by being made more vulnerable to the scrutiny of others, even if no one actually chooses to access their information.

The increased application of AI in medicine will greatly intensify the threat to privacy in the digital age both by driving the collection of more data and also by increasing the range of uses to which data may be put. The fact that AIs require millions of datapoints for their training provides an incentive to researchers to “hoover” up any and all available data. The sorts of data that may enable an AI to make predictions relevant to healthcare outcomes also include data generated outside of the healthcare system, such as histories of activity online or information about an individual’s lifestyle gathered by various apps or wearable devices. AI also dramatically increases the amount that can be gleaned from this data and thus the amount that people have at stake when it comes to the question of who can access it.

Optimists continue to hope that privacy can be maintained through technical measures. In particular, it might be thought that privacy can be preserved through the deidentification of medical data. But there are two problems here. Firstly, as we’ve noted, it is sometimes possible to determine health related information about an individual based on non-medical data. A salutary example of this occurred when Target revealed the pregnancy of a teenage girl to her family after having detected this on the basis of her purchasing history at the department store \citep{duhigg2013companies}. In order to maintain medical privacy, then, it would seem necessary for \textit{all} of our data to be deidentified as opposed to mere medical data, which is likely to be resisted by many of the companies that gather data insofar as their business models rely upon the ability to target advertising to individuals. But, secondly, even if all of the data that we generate as individuals were systematically deidentified, it is still possible for this data to be ‘reidentified’ once the amount of data reaches a certain – often surprisingly low – threshold \citep{gymrek2013identifying}.

It therefore seems likely that there is a trade-off between privacy and the potential healthcare benefits that might be realised through the use of AI. Some authors have argued that  we have a duty to forfeit our privacy and share our data in order to contribute to a ‘learning’ health care system that will be to the benefit of all \citep{cohen2018there}. This supposes that everyone has an equal likelihood of benefiting from the use of medical AI. However, as we shall see, the potential of AI to centralise political and institutional power, as well as the problem of algorithmic bias, means that the benefits of AI may very well be distributed unevenly.

\subsection{Surveillance}

Closely related – but not identical to – the threat to privacy is the danger of increased surveillance as a result of the capacities of AI. Privacy can be violated through the accidental release of data and even by the fact that people could access data, even if no one does. Surveillance consists in the deliberate gathering of information via the active scrutiny of populations. Surveillance may be morally problematic even where there is no expectation of privacy.

AI facilitates surveillance in at least three ways. First, as we’ve seen, AI makes it possible to gather more data and to gather new forms of data. By integrating information produced by sensors across multiple modalities, AI can produce data that is both richer and more fine-grained than ever before. Enthusiasm for big data, driven by AI, has led to researchers and corporations looking to the data generated by people’s online activities, including social media, and mobile phones for insights related to their health. Of course, once this data has been collected for purposes related to healthcare, it is also available for other investigations. Much of this data is now geo-tagged, making it possible to track people’s activities through time and space. Second, AI makes it much easier to work with large databases and to identify patterns within data. AI systems now regularly work with millions of records, in databases with many dimensions. Machine learning algorithms can identify correlations that are too subtle for human beings to observe directly. Third, AI can do all of this automatically, without direct human oversight. It can operate 24/7, often in real time, flagging relevant findings for human attention as required. 

The capacity of AI to enhance surveillance is a feature as much as a bug. For instance, many applications of AI in public health contexts or to identify iatrogenic harms rely on surveillance. Interestingly, physicians and healthcare workers are themselves likely one of the first targets of AI enhanced surveillance in order to monitor their performance \citep{dias2019using}. 

Nevertheless, it is clear that there are also significant ethical and political risks here. Part of what is problematic about surveillance is the loss of privacy that it involves, but this is not the whole of the matter. Surveillance, especially where licensed by the government, involves a fundamental change in the relations between organisations and individuals; between watcher and the watched. Individuals are interpolated — called into existence — as sources of risk. Everyone – or everything – is placed under suspicion. They are watched and measured, tagged with estimates of risk, and potentially targeted for intervention. The fact that individuals are thereby rendered vulnerable to the actions of the surveilling party is normatively significant even if that party never chooses to act on the basis of the information it has gathered \citep{pettit1997republicanism}. As we will discuss further below, the introduction of these powerful new tools of surveillance also tends to go hand in hand with the centralisation of power within institutions.

\subsection{Bias}

AI systems are only as good as the data upon which they are trained, and when this data is incomplete, unrepresentative, or misinterpreted, the results can be catastrophic. 

Outside of medicine, a number of high-profile cases have emerged where “bias” in the data used to train AIs have led to algorithms that produce discriminatory and/or offensive outputs. Typically, these biases disadvantage already marginalised and disadvantaged social groups.  One particularly high-profile instance of bias was Google’s image recognition software identifying African-American faces as those of gorillas \citep{barr2015google}. Similarly, online recruitment ads have been shown to present higher paid jobs to men and lower paid jobs to women \citep{spice2015questioning}. The implications of bias for machine learning can be especially pernicious if the outputs of the AI influence the nature of the data that is subsequently used to train it \citep{o2017weapons}. In such cases, bias may be self-reinforcing. For instance, the use of machine learning techniques in predictive policing, where police are sent to patrol areas that are identified by an AI – on the basis of historical data about where crimes occur, as likely sites of future crimes – has been linked to “increasingly disproportionate policing of historically over-policed communities” \citep[19]{lum2016predict}. In part this is because sending patrol cars to any location will result in an increase in reported crime given that it is the task of the police to identify and report crime.  By this mechanism, initial geographic variation in the reports of crime, often as a result of racist policing, may be rapidly amplified. 

The data used for medical research is not now, and is unlikely ever to be, free of bias. Sex, class, and gender, all influence who presents to hospital, with what conditions, and how they are treated. Physicians are, regrettably, not necessarily less susceptible to racism, sexism, or other forms of bigotry, than investigators in other sciences. There is, therefore, real danger that the use of AI in medicine might deepen existing inequities in health and healthcare along the lines of race, class, and sex. Particular social groups could be excluded from the benefits of medical AI or even actively harmed by medical AI systems. One striking example, of the how this might occur, which has already received some attention, relates to the use of AI for the diagnosis of skin cancers. The datasets that machine learning algorithms are being trained on for this purpose have tended to consist almost exclusively of fair-skinned individuals. Consequently, Adamson and Smith suggest that "[a]lthough there is enthusiasm about the expectation that ML [machine learning] technology could improve early detection rates, as it stands it is possible that the only populations to benefit are those with fair skin" \citep[1247]{adamson2018machine}.

Since biases often emerge from poor data collection and evaluation methods, it might be thought that engineers will be able to design around these problems through sufficiently careful curation and testing of data. But bias can emerge in an AI system even when the system is trained upon a high-quality datasets. Aggregation bias, for instance, results from the use of a one-size-fits-all model for populations with different statistical properties. It “can lead to a model that is not optimal for any group, or a model that is fit to the dominant population (if combined with representation bias)” \citep[5]{suresh2019framework} Suresh and Gutag give the example of clinical aid tools for diabetic patients: statistically relevant differences between diabetics with different ethnicities mean that, even if there is sufficient representation across ethnicities in a training dataset, a one-size-fits-all model will not serve the interests of each group equally or well.

Moreover, at a deeper level, the question of bias must be confronted rather than avoided. Data is always selected from a wider set of possible datapoints on the basis of assumptions, explicit or implicit, about the phenomenon it is being used to investigate. What counts as (problematic) “bias” is a methodological — and often an ethical — question. Similarly, given enough data an AI will find – or, as the example above suggests, obscure — multiple correlations, some of which we may wish to factor into our analysis of the phenomenon and some of which we way wish to reject as artefacts or on ethical or political grounds. Concerns about bias in AI should serve as prompts for ongoing conversations about the basis on which we wish to make these decisions rather than be thought to constitute a reason to abjure AI.

\subsection{Explainability}

A troubling property of some AI systems is that they may function as ‘black boxes’. This is especially the case where an AI utilises the complex computational architecture of neural networks. In deep learning, neural networks have multiple ‘hidden’ layers that each contain a large number of artificial neurons. Each of these neurons hold a particular statistical weight or ‘bias’ that influences the final output of the system. During training, biases are commonly assigned at random and then optimised autonomously through a technique called backpropagation, wherein a particular backpropagation algorithm moves back through the layers of the network in reverse order, adjusting the biases of individual neurons in order to optimise the overall performance of the model. The reasoning behind each of these innumerable adjustments is not accessible, severely restricting our ability to understand or offer an explanation for the systems outputs once it has been trained. Perversely, the most accurate and useful AIs are often the least explainable \citep{burrell2016machine}. 

A number of authors have worried that the opacity of AI systems limits the capacity of designers to identify and mitigate risks to patients \citep{cabitza2017unintended, terrasse2019social, watson2019clinical}. This worry does, however, need to be placed in perspective. Lack of explainability is already widely accepted across various domains of medical practice. London, for example, notes that "modern clinicians prescribed aspirin as an analgesic for nearly a century without understanding the mechanism through which it works. Lithium has been used as a mood stabilizer for half a century, yet why it works remains uncertain" \citep[17]{london2019artificial}. If we can be justified in prescribing a drug without being able to explain how it works, or why it produces adverse outcomes in some patients but not others, then it seems we could be justified in relying on an AI that was generally reliable even where we don’t understand how it manages to be so. To the extent that we are solely concerned with medical outcomes, narrowly conceived, then the difficulty in explaining the internal functioning of AI does not distinguish them from other lacunae in medicine.

Yet the practice of medicine involves more than just the cure of disease or illness – it also involves relations between persons. Explanation and understanding have become increasingly important to patients, as evidenced by the turn away from paternalism towards patient autonomy in medical ethics and practice over recent decades. Patients want to make their own choices about their health, or at least share the decision with their doctor. In this context, the opacity of AI does indeed appear to be problematic: it deprives patients of the opportunity to receive answers about key questions related to their treatment. As we shall discuss further below, it also makes it difficult to identify the value judgments that have been made in the course of reaching a treatment decision or to check that the decision is in accordance with the patient’s own values. Moreover, in many contexts, medical decisions raise questions of justice, about the allocation of resources amongst persons. A Kantian notion of respect for autonomous agents requires that we can provide reasons to justify our treatment of other people. Ideally, a condition of a purported reason being a reason is that it is potentially something that the other party should accept — or at least recognise as normatively relevant to the situation. Arguably, the deliberations of black boxes can’t constitute reasons in this sense, as we are unable to follow or evaluate them. If people want to be able to make medical decisions in accordance with their values, or if people are going to be treated differently because of the deliberations of a machine, then, we have reason to prefer systems that enable us to track the reasons for the conclusions that they reach.

\subsection{Trust}

The difficulty in explaining the deliberations of AI may also have implications for the future of trust in medicine and healthcare systems. Trust in one’s clinician has important advantages in any medical encounter. It allows one to feel comfortable revealing personal and sensitive information to them, to feel confident in their judgment and advice, and to comfortably depend upon them in times of ill health and vulnerability. For these reasons, trust in one’s clinician has been shown to have a positive correlation with improved self-reported health \citep{hall2001trust}. Yet the likely intrusion of AI into the clinical encounter in the near- to mid-future has the potential to hinder the development and maintenance of patient trust \citep{vayena2018machine}, possibly jeopardising some of the benefits that medical AI is expected to deliver. The problem of explainability, addressed in the previous section, is likely to have significant influence here. “If doctors don’t understand why the algorithm made a diagnosis,” as Watson and colleagues observe, “then why should patients trust the recommended course of treatment?” \citep[2]{watson2019clinical}. Indeed, the more that clinicians and patients come to rely upon the use of medical AI, the more that relations of trust may shift away from human clinicians toward the AI systems themselves. 

Patients are not the only stakeholders for which trust in AI is problematic. Clinicians, too, face challenges here, since they will likely be expected to mediate between patients and AI systems \citep{verghese2018computer}. It is crucial for patient safety, then, that they understand when it is appropriate and inappropriate to place their trust in these systems. The problem, however, is that the automation of tasks often leads people to both over-trust and under-trust these systems in different contexts, with potentially disastrous implications.

Automation bias, for instance, is one example of over-trusting, which occurs when people rely too heavily upon systems that they have observed over time to be generally accurate, leading to otherwise avoidable error. The Therac-25 disaster is one example of this phenomenon in medicine. Therac-25 was a computer-controlled radiation therapy machine that inadvertently gave radiation overdoses to six patients, resulting in serious illnesses and deaths. Troubling clinical observations during these overdoses were disregarded because the operators of Therac-25 came to mistakenly trust the machine over their own expertise \citep{ash2004some}. The use of AI in medicine has significant potential to lead to similar instances of over-trust. 

\textit{Under}-trusting AI might also prove problematic. An existing example of under-trust in clinical settings is seen in the phenomenon of alert fatigue, where hospital staff come to disregard computerised alerts because of their interminable frequency and clinical irrelevancy. Under-trust of this sort can lead to avoidable patient harm. Wachter, for instance, has detailed an instance wherein a patient received a 37-fold overdose of antibiotics, despite a number of computerised warnings of the error which were ignored because of under-trust in the systems’ alerts \citep{wachter2017digital}. AI in medicine could have similar effects if clinicians come to distrust their outputs or suggestions if these outputs are not communicated thoughtfully and effectively, which has significant potential to lead to patient harm. 

\subsection{De-skilling}

The problem of automation bias is exacerbated by de-skilling. Changes in the skill-sets of doctors are nothing new and are indeed a natural consequence of progress in medicine. Physicians lose skills when they rely on machines to perform tasks that they use to perform unaided or — perhaps more often nowadays — when they rely upon new machines to perform tasks that they use to perform with the assistance of older machines; as these tools improve the skills required to use previous generations thereof disappear. Deskilling may arise at three different loci. Individual physicians may gradually become unable to perform tasks that they were once capable of performing, owing to lack of practice as a result of a new technology rendering the skill redundant. Individual physicians may no longer learn skills that were once taught to doctors as new technologies render the old skills redundant. Finally, the profession as a whole may lose a skill if no one remembers how to perform a task that doctors use to perform before the new technologies arrived. The prospect of deskilling as a result of AI, though, seems especially unsettling because people are now talking about AIs outperforming humans in roles that have previously been thought to be the very centre of the practice of medicine, including diagnosis and prescription.

One reason to be concerned about such deskilling is pragmatic \citep{carr2015glass}. We may worry about the implication of a loss of skill for an individual’s — or perhaps the profession’s — ability to achieve some goal. Given that skills are eroded only when regular exercise of the skill becomes unnecessary, this effectively means a concern about what might happen if the technology fails. In the future, will doctors still be capable of diagnosing and treating people in those — hopefully — rare situations in which AI is not available, perhaps due to a failure of the power supply, or because of a system “crashing”? The force of this concern, then, will depend on a number of considerations. How likely is it that AI might fail and for how long? How important is the procedure that the AI facilitated and how urgent is it? What alternatives exist if doctors have lost the skills they would have previously relied upon in these circumstances? Interestingly, the balance of these considerations is likely to alter over time. The first generation of doctors to work with AI are less likely to be de-skilled than our subsequent generations. However, presumably AI is likely to become increasingly reliable over time, so it will matter less if doctors don’t have the skills of previous generations.

However, a second set of worries about deskilling comes to the fore when the exercise of a skill is valuable for its own sake or because it is implicated in some other inherent good \citep{carr2015glass,danaher2018toward}. It is plausible to think that the exercise of some skills – those constitutive of the virtues — is essential to having a flourishing human life \citep{vallor2015moral}. For instance, a person who never exercises practical wisdom — reasoned about their own ends — because they possessed an AI that deliberated on their behalf would not thereby be made any better off.

How one should relate to the prospect of the deskilling of physicians as a result of the increasing presence of AI in medicine, then, depends upon whether one thinks of medicine as being solely instrumentally valuable in promoting health or hold that there are aspects of the practice of medicine — including, as we discuss further below, the opportunity it presents to demonstrate care — which are inherently valuable.

\subsection{Fragility}

The pragmatic concerns about the impact of deskilling mentioned above are especially pressing given that the adoption of AI may often introduce a single point of failure in medical care: in some contexts, nearly everyone will be relying on the same system. Where AIs can perform tasks that were previously performed by human beings there is no reason why they cannot replace every human being who was performing that task. That is, a single AI could become responsible for all detection of skin cancer or analysis of chest x-rays, et cetera.

Indeed, there is at least one ethical reason why this should occur, and two pragmatic/political reasons why it is likely to occur. 

The ethical case for AI monopoly rest on the duty of nonmaleficence (or alternatively on beneficence). Where there are multiple AIs that can perform a given task their performance is likely to vary. Failure to employ the best system will harm patients and so every institution will be under a moral obligation to adopt the best AI. The pragmatics of the marketing of AI, which will inevitably emphasise its performance using standard metrics, will also make it politically difficult to do anything else. No institution wants to be seen to be offering an inferior service, let alone to be seen to be putting the lives of their patients at risk.

These arguments might be less compelling were there only to be minor differences in the performance of competing AIs. However, there are reasons to anticipate that one AI will often come to offer clearly superior performance as compared to its competitors at a particular task. The role played by big data in AI means that early competitive advantage, especially higher market share, is likely to lead to market dominance. The more users a system has, the more data it will have access to… and thus the more it can learn on the basis of this data. This dynamic will encourage effective monopoly with regards to the provision of particular services by AI.

Should a medical AI malfunction, then, the consequences are likely to be disastrous. Where an incompetent or malicious human physician might harm dozens of patients, mistakes made by an AI may affect hundreds of thousands of patients. Indeed, even if an AI is not the sole provider of the service it provides, it is likely to be involved in the care of many more patients than any human being could be. The risks here are exacerbated by deskilling and automation bias, both of which make it less likely that physicians will detect problems at an early enough stage to avert widespread harms.

The dangers posed by a single point of failure should prompt healthcare providers to employ more than one system in order that each system might serve as a backup if the others fail. Unfortunately, this may not always be possible, will usually be expensive where it is possible, and raises ethical questions of its own. It may not be possible because, for the reasons noted above, particular AI systems may outcompete all others at some particular task so as to effectively establish a monopoly. Even when multiple providers exist, employing more than one will usually be very expensive because a significant portion of the cost of AI is generated by the need to integrate the AI into a hospitals’ (or other institutions’) IT systems, workflow practices, and electronic medical records (where these exist). Indeed, often an institution’s practices and IT systems will need to be modified to suit the demands of the AI. Having to do this for multiple systems greatly complexify the task and increases the expense of performing it. One of the ethical issues raised by the desire to sustain a fallback system, then, relates to the opportunity costs associated with this expense: the funds required to support this might instead be used to benefit patients more directly. In theory, it should be possible to evaluate the cost-effectiveness of deliberately diversifying when it comes to AI in key roles, which will depend to a large degree on the probability of either system failing, the consequences of failure in the absence of another AI alternative, and the cost of introducing a second system. However, in practice, it will be extremely difficult to estimate either the risk of failure of any given AI system or the costs of introducing a second system.

\subsection{Power}

Thinking about a scenario in which there is a single provider of a key medical service also highlights the power possessed by whichever corporation designs this AI. There is a real risk of vested interests here, exacerbated by the potential such power allows for manufacturers of the AI to hold healthcare systems — and patients — to ransom when it comes to the pricing of the service and/or its future development.

The introduction of AI will also have other implications for the distribution of power within healthcare systems, which are worth highlighting. We have already noted the way in which AI facilitates surveillance and thus empowers governments relative to citizens and corporations and institutions relative to individuals. Deskilling of physicians, should it occur, will reduce their social standing relative to other professions and their bargaining position within healthcare institutions. By contrast, the more institutions rely on AI, the more power computer scientists and IT departments will accrue within them. Although it is difficult to say too much about them in the abstract, these shifts in power within institutions are, we submit, one of the most likely and important impacts of the advent of medical AI.

\subsection{Responsibility}

The introduction of AI into medicine will also have implications for the allocation of responsibility for treatment decisions and adverse outcomes. Who should be held responsible when things go wrong as the result of a decision that depended crucially on the output of an AI? The manufacturer of the AI? The designer of the AI? The institution that purchased the AI? The physician? Or — most controversially — the machine itself? 

Discussions of these questions often stumble over the difficulty of providing a clear account of the autonomy, and the agency, of AI \citep{johnson2013artefactual}. It is tempting to think of AI, especially AI involving machine learning, as creating a “responsibility gap”, such that it becomes impossible to allocate responsibility for any of the human parties, who could not have known precisely how the AI was going to act \citep{matthias2004responsibility}. Allocating responsibility to the machine itself is problematic because concepts like guilt, shame, and punishment, which are essential to our thinking about responsibility have little purchase when it comes to machines \citep{sparrow2007killer}. 

The problem with this line of thought is that mere uncertainty is no barrier to the allocation of responsibility. Uncertainty — about the precise aetiology of a patient’s symptoms, about whether a patient will respond to a particular drug, or about the future progress of the disease – is, after all, endemic to medicine and poses no especial difficulty when it comes to the attribution of moral or legal responsibility. Doctors must make decisions on the basis of the information available to them, and we assess their responsibility for adverse outcomes accordingly. While the cause of a patient’s death might be that a cancer did not respond to the treatment provided, the responsibility for the treatment decision remains the doctors and if anyone should be held morally responsible for the death this will depend on whether the decision was justified given the information available to the doctor about the likelihood that the cancer would respond. Similarly, even if doctors don’t know precisely how an AI will perform in relation to the treatment of a particular patient, this uncertainty doesn’t prevent us from assessing whether their decision to rely on it was reasonable or not and therefore whether they should be held responsible for the outcome of the course of treatment suggested by the AI. A “responsibility gap” would only emerge if the AI had agency — or at least a form of pseudo-agency — sufficient to imply that the machine might sometimes be morally, and not just causally, responsible for the outcome of acting on its advice. While one cannot rule out the possibility of machines developing such agency in the future, none of the AI systems currently on the horizon — not even those involving deep learning — are plausibly thought of as moral, or even pseudo-moral agents.

If we think, instead, about an AI being reliable, or fallible, in the same way that a cancer medication is reliable, or fallible, then it is possible to make progress. Adopting such a deflationary account of the agency of AI allows to us to see how a familiar set of intuitions and principles can guide us in allocating responsibility for the outcomes of medical treatment involving AI along the same lines that we distribute it for outcomes involving any other complex technology. The design and performance of the AI is the responsibility of the designer. Responsibility for the use of the AI will usually be shared between the physician and the healthcare institution within which the treatment is provided. Responsibility for acting on the basis of the output of the AI will rest with the treating physician. 

When something goes wrong, we must ask whether any of these parties failed in their obligations and attribute responsibility accordingly. In some cases, it may well be appropriate to conclude that \textit{none} of these parties have done anything wrong. This may include cases where the AI behaved in an unanticipated fashion. Yet this no more involves a responsibility gap than when a patient has an allergic reaction to a drug that could not reasonably have been anticipated. As long as the machine is sufficiently reliable at the task it was expected to perform there may be no wrongdoing involved even in those cases where it fails.\footnote{The fact that AIs involving machine learning may behave in ways that are difficult to predict \textit{is} relevant to the responsibilities of designers and physicians, with the former having an obligation to try to reduce the uncertainty about the performance of the machine in any given case, and the latter having an obligation to take this fact into account when deciding whether to defer to an AI. Note, however, that the manufacturers and prescribers of drugs have the very same obligations.}

There is, of course a second set of issues about how AI will impact on responsibility, which concerns the way the development of AI might change the duties of doctors and other healthcare providers. We have already suggested that institutions might come to be held to be under an obligation to purchase only the best AI available. Similarly, adopting AI is unlikely to remain optional for physicians very long: it will eventually become morally required. Whenever the use of machine brings about better outcomes for patients than human beings employed in the same task it will be obligatory to defer to the machines.

\subsection{Framing}

A key insight from the philosophy of technology is that tools are never just tools. They are never “neutral”. Instead, they shape our ends \citep{winner2020whale}. Tools have “affordances” – they make it easier to do some things as compared to others and by virtue of this fact they “frame” problems. Indeed, as we have already noted, by altering expectations, technologies may effectively require those who have access to them to adopt them and use them in the pursuit of particular ends. 

Like any technology, then, AI will have values built in. Some of these values will reflect choices made by their designers: some of them may reflect the society from which the data used to train the AI was sourced. However, some may be implicit in the very way AI frames the problems it then works to address. 

To a person with a hammer, everything looks like a nail. To a healthcare institution with an AI, everything may look like data. We believe that there is a real danger that taking up the tool of AI will subtly reshape the goals and nature of the practise of medicine: AI frames the problem of restoring and promoting health as a problem of \textit{information}. To secure health we need more information: if we have more information, we will have more health. \cite{verghese2008culture} has observed how the advent of the computer, electronic medical records, and sophisticated medical texts has redirected the attention of physicians away from the body of the patient towards the patient’s data. AI risks a further level of abstraction away from the particular patient’s data towards data “in the cloud”, with individual patients and their data appearing primarily as data points. Of course, the practice of medicine has always involved the pursuit of understanding, both about the origins of a particular patient’s health problems and about the functioning of the human body and the nature and causes of its diseases. But as Verghese notes, historically, the pursuit of such understanding has involved “knowing how”– how to learn about the patient’s condition, as well as to cure it, by the exercise of skill practiced upon their body — as much as “knowing that”. The skills required for medicine in the future may be increasingly oriented towards data and be possessed mostly by data scientists rather than physicians.

\subsection{Care}

Human contact, attention, and empathy have therapeutic value. They also play a more foundational role in the practice of medicine. Doctors are not (just) mechanics of the human body: medicine is fundamentally a caring profession.

Although, as discussed above, some authors believe that AI has the potential to enhance care, medical AI might also be thought to constitute a \textit{threat} to care as more and more medicine is delivered by machines. Machines can’t care \citep{sparrow2006hands}, so the more medical roles they take on the scarcer opportunities to demonstrate and experience care may become. 

It is already the case that medical practice has come to be dominated by data and test results at the cost of the patient-physician relationship \citep{verghese2008culture}. Increasingly, care of the sort that is expressed in touch, gesture and gaze in the course of ministrations to the patient’s physical body is the province of nurses and allied health professionals. Insofar as these professions tend, in many parts of the world, to be dominated by women, the provision of this care is correspondingly gendered.\footnote{Most care actually occurs outside of the formal practice of medicine, in the home, and in that context is overwhelmingly provided by women.} The future of care will therefore depend in part on whether AI comes to replace the role of doctor or the role of nurse. 

Because it is much easier for AI to deal with “data” online than it is for machines to function alongside of human beings in the physical world there is every reason to expect that the work of doctors will be taken over by AI long before the work of nurses. One important reason to be concerned about this prospect relates to the role played by empathic human contact in the professional self-understanding of physicians – and consequently in their level of satisfaction with their work \citep{truog2019slide}. Another is the role of care in motivating behavioural change. Advice from a computer, or from a physician who is perceived to be little more than the mouthpiece of a computer, may not be sufficient to get people to follow a course of treatment or address the lifestyle factors that are implicated in their health problems. If people don’t feel that their doctor really cares about them, they may be less concerned with what their doctor thinks about them and thus to take their doctor’s advice. The extent to which people change their behaviour on the basis of advice provided by machines is, of course an empirical matter, about which we now have some data relating to the use of health and lifestyle apps: while this data is mixed, it has to be said that it does not inspire much confidence \citep{arigo2019history, finkelstein2016effectiveness, jakicic2016effect, mckay2018evaluating}.

There is undoubtedly a risk of anachronism in worrying about the role of care in a future in which AI plays a greater role in medicine. If offered the choice between human-directed medicine, with lots of human contact, emotional support, and care, but with uncertain outcomes, and treatment by cold uncaring machines, which would cure them of their ills, patients might not unreasonably opt for the latter. In reality, for the reasons we have rehearsed here, patients are unlikely ever to be faced with such a stark choice: in practice, medicine that involves care is likely to be more effective. Given the institutional – and ethical – imperatives to embrace the use of AI in medicine, surveyed above, philosophers, bioethicists, and others concerned with the future of patient-centred medicine would be well advised to prioritise the development of a robust defence of the value and role of care in medicine in order to ensure that patients aren’t asked to confront this false dichotomy regardless.

\subsection{Values}

Finally, it will also be important to think about how AI might impact on our capacity to reason about, and defend, values more generally. As \cite{habermas1985theory}, as well as other members of the Frankfurt School \citep{horkheimer2002dialectic}, have argued at length, technical (or “instrumental”) rationality tends to crowd out reasoning about ends. AI promises to be an immensely powerful instrument. Yet medicine often requires us to think about ends. Especially at the beginning and end of life, questions about the nature of human flourishing, and/or about how to balance respect for autonomy versus a concern for the best interests of the patient, loom large. As we’ve noted, AI is likely to have some values already built in, but the opaque nature of many of these systems will make it especially difficult to allow for these values when it comes to making a decision based on the output of an AI. More generally, as we also observed earlier, where the internal operations of AI systems are opaque, it may be difficult to assess how the results of their prognostications should be taken into account in our thinking when we are thinking about ends. There is a significant danger that the power of AI to solve problems about \textit{how} to do things will lead to doctors and patients spending less time deliberating about \textit{what} to do.

Conversely, when questions about ends do arise, AI will have little to offer. How much should I value the opportunity to have children in years to come? Should I pursue longevity at the risk of losing my dignity? Should I turn off my father’s ventilator? These aren’t things that machines will be able to reason about for the foreseeable future. Indeed, it is hard to imagine how machines could ever offer anything in relation to such questions given that machines can’t stand behind their claims in the way that people must before we should take their moral advice seriously \citep{gaita2004good}.

The role played by deliberation about ends in medicine therefore offers some comfort to those who worry about the possibility that AI’s will replace doctors entirely. Just how much comfort it offers depends upon how plausible we think it is that physicians will continue to be able to advise, or assist, patients in their deliberations about their values, and the implication of their values for their medical care, when more and more of the routine practice of medicine is directed by machines.

\section{Conclusion}

Artificial intelligence has much to offer patients, doctors, and healthcare systems. Inevitably, with potential benefits it also brings risks. We have highlighted the potential for AI to facilitate medical research, more accurate diagnosis, and more efficient medical administration. We have also drawn attention to the likelihood that AI will threaten patient privacy and facilitate surveillance. Biased data may jeopardise the benefits of AI and lack of explainability should sometimes – but not always – reduce the extent to which we are willing to rely on it. How much we are likely to be willing to trust AI, as well as how much we should, remain open questions. De-skilling of physicians as a result of AI is a significant risk: over reliance on AI may also render healthcare systems more fragile. We have highlighted the prospect that the introduction of AI into healthcare may empower some at the expense of others. We have also argued that the use of AI is less problematic for the allocation of responsibility than is often suggested. Finally, we have raised concerns about the ways in which AI might reframe medicine, impact on care, and discourage important arguments about values in medical decision making. 

A thorough investigation of any – let alone all – of these matters must needs draw on the combined expertise of physicians, engineers, data scientist, economists, political scientists, sociologists, and science and technology scholars, as well as philosophers and bioethicists. Pending the results of such a larger study, we hope this initial survey has at least identified some of the key questions as well as promising lines for future inquiries. We also hope that might be of some use to physicians and policy makers who are already grappling with the implications of AI and thus to realise the benefits of AI in medicine.

\section*{Acknowledgments}

The authors would like to thank Prof Ruiping Fan, as well as the commentators, for their patience over the course of the drafting of this manuscript. Thanks are also due to Prof Fan for the invitation to present an earlier version at the "International Workshop on 'Ethics of Biomedical Technology and Artificial Intelligence'" at the City University of Hong Kong. Dr Sparrow would also like to thank Dr Derrick Au for a previous invitation to Hong Kong to present on AI in medicine and for conversations and discussion at the time. 

\bibliography{main}
\end{document}